\documentclass[manuscript]{aastex}

\shorttitle{Methods for Distinguishing Rotation and Reversal}
\shortauthors{Z. X. Liang \& Y. Liang}

\begin{document}

\title{Six Methods for Distinguishing Rotation and Reversal \\ in a Stellar Magnetic Field}

\author{Zhu-Xing Liang\altaffilmark{}}
\affil{KPT lib, Shuixiehuadu18-4-102, Zhufengdajie, Shijiazhuang,
China 050035}\email{zx.liang55@gmail.com}

\author{Yi Liang\altaffilmark{}}
\affil{College of Physics, Jilin University, Changchun, China
130031} \email{jluliangyi@gmail.com}

\shortauthors{Liang \& Liang}

\begin{abstract}
    Periodic changes in a stellar magnetic field can be explained in two
   ways: the oblique rotator model or the solar cycle model.
   Although many papers favor the oblique rotator model,
   there has not been enough evidence to rule out the solar cycle model definitively.
   This paper presents five new observations (together with a old one)
   that can be carried out to distinguish magnetic field rotation and magnetic field reversal.

\end{abstract}

\keywords{stars: activity -- stars: magnetic fields -- stars:
rotation}

\section{Introduction}

Periodic changes in a stellar magnetic field can generally be
explained in two ways: the oblique rotator model assumes that the
magnetic field rotates, while the solar cycle model assumes that the
magnetic field reverses. Although many papers favor the oblique
rotator model, to date there has been no reason to discard the solar
cycle model definitively. \citet{B51, B58} once discussed the
advantages and disadvantages of these two models with respect to
interpreting data from typical $\alpha$-variables. While the
available data favored solar cycle (reversal) models, Babcock did
not arrive at a definitive conclusion.

The intensity and polarity of a stellar magnetic field have
traditionally been used to compare models, but these data provide
only a limited view of the magnetic field behavior. If other types
of data could be obtained, the choice of model would be more
convincing.

This paper discusses five new methods and one old method of
determining the reason for periodic changes in a stellar magnetic
field. All the new methods make use of additional data such as the
line-of-sight magnetic field, the transverse magnetic field, the
polarization angle, and radio-pulse radiation. Used together, the
true nature of a stellar magnetic field change can be ascertained
correctly.

\section{Six Methods for Distinguishing Rotation and
Reversal in a Stellar Magnetic Field}

\subsection{Observing Polarity Changes in a Population}
The first method has been used before. If the magnetic field is
reversing, its polarity must change twice per cycle. If the magnetic
field rotates, then its intensity must undulates periodically but
its polarity may not reverse. The necessary conditions for observing
a polarity reversal are: 1) the magnetic inclination (the angle
between the magnetic axis and the spin axis) is not equal to zero;
2) the spin inclination (between the spin axis and the line of
sight) is not equal to zero; and 3) the sum of magnetic inclination
and spin inclination is greater than 90 degrees. In other words, if
the magnetic inclination and/or spin inclination is too small, the
polarity will not appear to reverse. If all objects in a large
sample exhibit polarity reversal, the oblique rotator model can be
ruled out. \citet{B58} used this method to analyze data on typical
$\alpha$-variables, resulting in forceful support for the reversal
model.

Although the later statistics \citep{B82} indicated that only
approximately two-thirds of Ap/Bp stars with periodically varying
magnetic field measurements are known to exhibit polarity reversal,
the difference between this result and Babcock's maybe come from the
illegibility of weak signals. In order to have a strong persuasion,
the samples with blurrier data should first be removed and then only
the reliable data be used as evidences. If all the samples with good
periodicity and high signal-noise ratio exhibit polarity reversal,
especially, if they all exhibit a fair degree of symmetry about the
\emph{\textbf{H}} = 0 axis in their reversals, it is sure that the
model of magnetic field reversal is correct.

\subsection{Phase Relationship of Two Field Components}
\begin{figure}
\epsscale{.80} \plotone{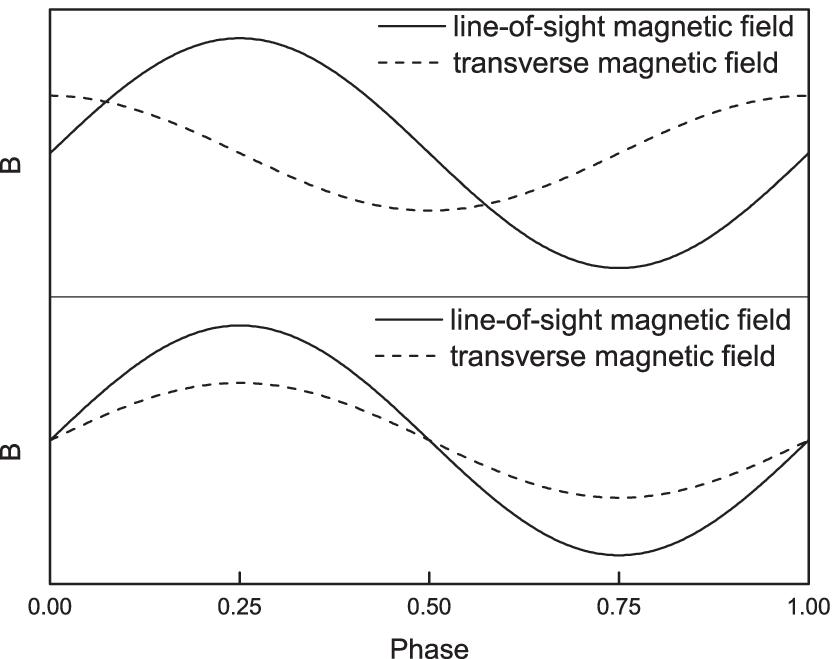} \caption{Relationship between
line-of-sight and transverse
   magnetic field components in the two models. The upper panel
   shows a rotating case, where the phase difference is 90 degrees.
   The lower panel shows a reversal process, where the two components
    have the same phase.\label{Fig1}}
\end{figure}

The second method compares the line-of-sight and transverse magnetic
field phases. When a magnetic field reverses, the two components
should change at the same time. If the magnetic field is rotating,
however, there will be a 90 degree phase difference between the two
components. This can be observed through the Zeeman Effect: the
magnetic field both splits and polarizes spectral lines. The
line-of-sight magnetic field will induce circular polarization in
the spectral line, while the transverse magnetic field will induce
linear polarization. The two components of the magnetic field can be
determined simultaneously by measuring changes in the line width. As
shown in the lower panel of Fig.~\ref{Fig1}, it is obvious if the
two components are in phase; in this case the reversal model must be
selected. If the phase difference of the two curves is about 90
degrees, as shown in the upper panel of Fig.~\ref{Fig1}, the oblique
rotator model is more appropriate.

\subsection{Relationship between Polarization Angle and Magnetic Field Intensity}
The transverse magnetic field can cause Zeeman splitting, in which
case the wings of a spectral line are mainly composed of linearly
polarized light. The polarization angle of the spectral line can
also help determine whether the magnetic field rotates or reverses.
If the magnetic field is rotating, its polarization angle will swing
back and forth between two values. If the magnetic field reverses,
its polarization angle will be fixed. Note that in both situations,
Faraday rotation has been ignored.

If we consider Faraday rotation, then magnetic field reversal can
also induce periodic changes in the polarization angle. This
oscillation is driven by changes in the magnetic field intensity. In
this case, the nature of the magnetic field can still be determined.
At the moment corresponding to the polarization angle's middle value
(halfway between its extrema), the intensity of the line-of-sight
magnetic field should reach its maximum under the rotation model. If
the intensity of the line-of-sight magnetic field instead reached
its minimum value, the magnetic field must have reversed. In fact,
this method is based on the same principle as that described in
Section 2.2. The only difference is that the intensity of the
transverse magnetic field is inferred from the polarization rather
than being measured directly.

\subsection{Phase Relationship between Radio Pulses and the Magnetic Field}
Pulsed radiation was once considered a unique behavior of neutron
stars, related to their extreme magnetic fields. Recent detections
of pulsed radiation from other stars such as the ultracool dwarf
TVLM 513-46546 (hereafter TVLM 513; Hallinan et al. 2006, 2007) and
the magnetic chemically peculiar star CU Virginis (HD124224;
Trigilio et al. 2000, 2008), however, have provided new insights
into the behavior of stellar magnetic fields. \citet{L07} put forth
a magnetic oscillation model (hereafter the MO model) which relates
pulsed radiation to the solar cycle model. Thus, observational
support for the MO model would significantly enhance the solar cycle
model too.

A fundamental assertion of the MO model is that radio pulses (at
least narrow pulses) always appear close to the zero point of
magnetic field intensity. The observations of \citet{T00} indicate
that this is the case for CU Virginis. While this is just one star,
we think it unlikely that the result is pure coincidence. We
therefore propose further investigation of the phase relationship
between magnetic field intensity and radiation, in order to
determine whether this correlation is a universal phenomenon.

\begin{figure}
\epsscale{.80} \plotone{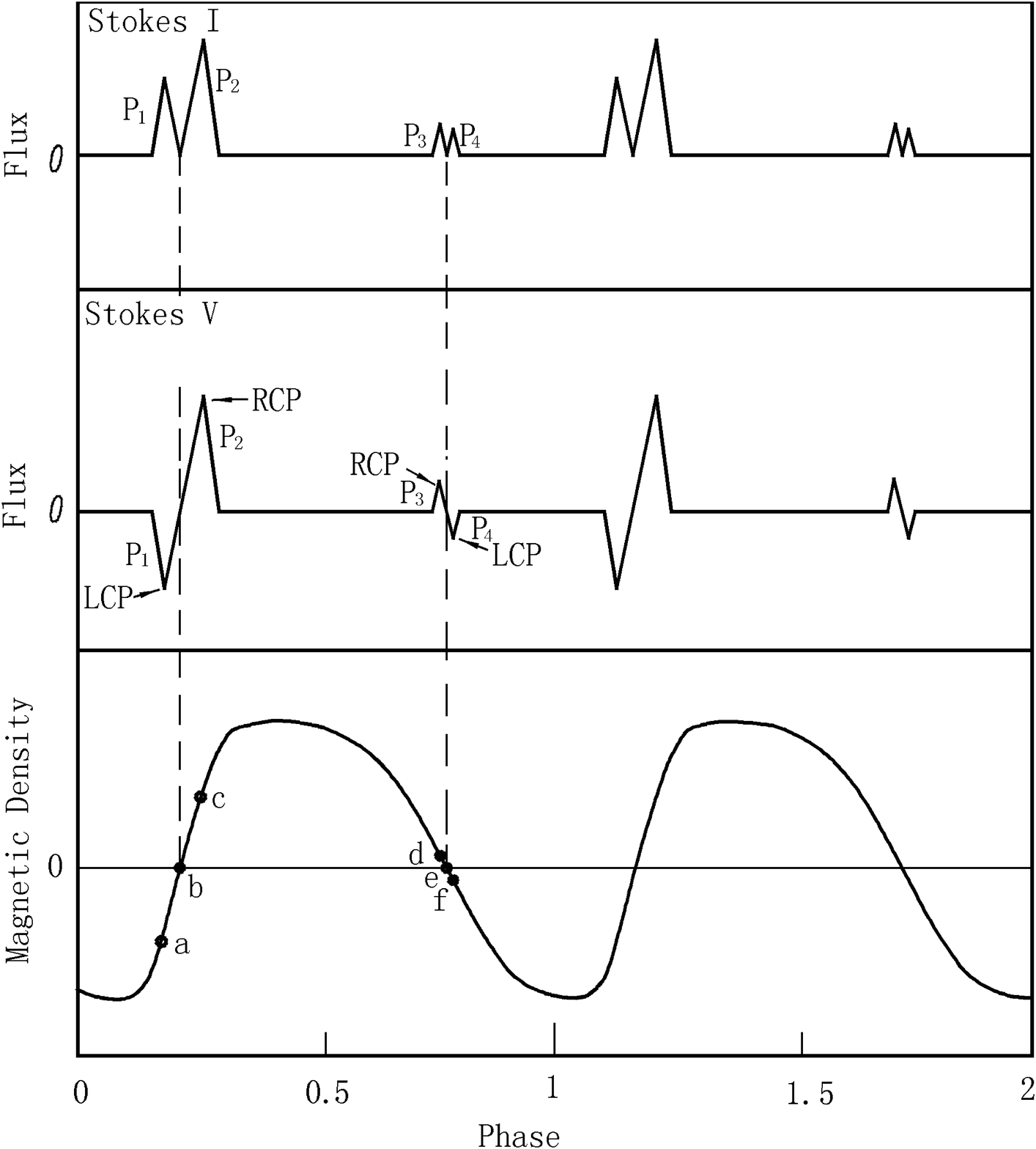} \caption{Relationship between
magnetic field intensity and radio pulses. Because the magnetic
field intensity is zero at point (b), radiation produced during the
(a)-(c) period will be split into the two sub-pulses $P_1$ and
$P_2$. In the same way, the pulse produced during the (d)-(f) period
will be split into sub-pulses $P_3$ and $P_4$. Left- and
right-handed circular polarizations are marked by LCP and RCP. When
the line of sight aligns with the rotation axis, the polarization of
$P_1$ will be opposite to that of $P_2$. Similarly, the polarization
of $P_3$ will be opposite to that of $P_4$. Because their magnetic
field directions are the same, however, the polarization of $P_2$ is
identical to that of $P_3$.\label{Fig2}}
\end{figure}

Another claim of the MO model is that like the sun, all stars
reverse their magnetic field periodically. Only the frequency of
this reversal is different. Near points (b) and (e) in
Fig.~\ref{Fig2}, where the magnetic field reverses direction, the
rate of change in magnetic flux is greatest and a very strong
electromotive force is induced. This causes charged particles to
circulate at very high speed, producing a pulse of gyroscopic or
synchrotron radiation. The observed radiation is pulsed because the
magnetic field only changes quickly over short periods around point
(b) and point (e). Furthermore, since the magnetic field intensity
is zero at points (b) and (e), the radiation drops to zero halfway
through the pulse. The radiation profile is therefore composed of
four pulses per period: $P_1$, $P_2$, $P_3$ and $P_4$ depicted in
Fig.~\ref{Fig2}.

The situation described above has been simplified, of course. For
one thing, the magnetic field intensity may not follow a sinusoidal
pattern. Thus, while the magnetic field always goes through zero
twice, its derivative might be different for each pulse. The peak
speeds attained by charged particles in the magnetic field will vary
accordingly, as will the amplitudes of pulses $P_1$, $P_2$, $P_3$
and $P_4$. It is possible for weaker pulses to be overwhelmed by
noise and escape detection. It is therefore normal to receive only
one, two, or three pulses per period rather than all four.

The MO model assumes that the magnetic axis of a star always aligns
with its spin axis. \citet{L07} pointed out that if the line of
sight aligns with this axis, the radiation should exhibit circular
polarization. In this case the order of pulse polarizations,
assuming all pulses are detected, should be LCP $\rightarrow$ RCP
$\rightarrow$ RCP $\rightarrow$ LCP $\rightarrow$ LCP as shown in
the middle panel of Fig.~\ref{Fig2}.

\begin{figure}
\epsscale{.80} \plotone{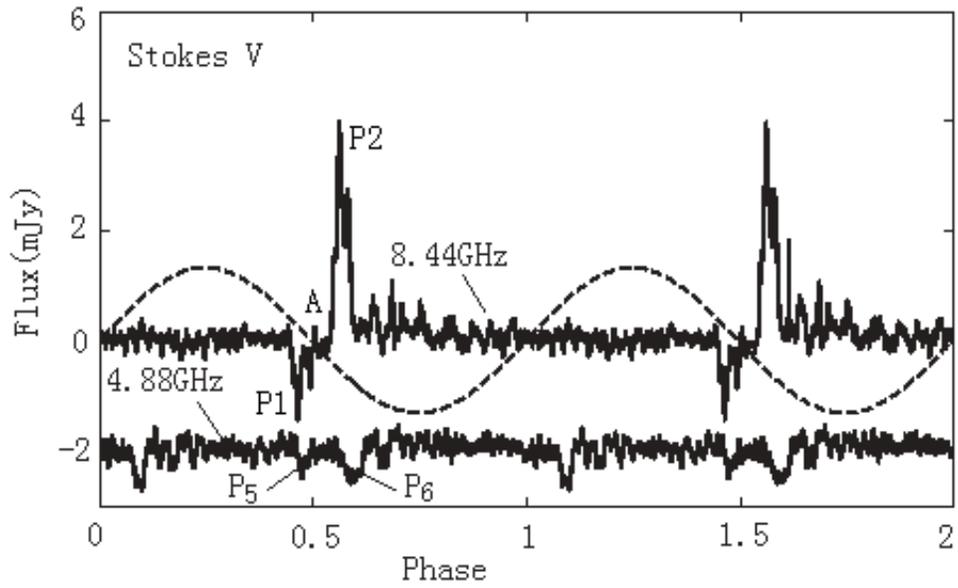} \caption{Radio light curves
detected from TVLM 513, reproduced from \citet{H07}, where the right
circular polarization is represented by positive values. The dashed
line represents the phase of the magnetic field, as predicted by the
MO model. The magnetic field null is at point A, between $P_1$ and
$P_2$.\label{Fig3}}
\end{figure}

The radio light curves shown in Fig.~\ref{Fig3} are reproduced from
\citet{H07} who detected radio emission from TVLM 513. According to
the MO model, 100\% circular polarization indicates that both spin
axis and magnetic axis are aligned with the line of sight. In
Fig.~\ref{Fig3}, pulses $P_5$ and $P_6$ on the corresponding
polarization curve are both left-handed. This result does not accord
with the MO model shown in Fig.~\ref{Fig2}. It may be that something
went wrong in processing these data. In another observation of the
same star \citep{H06}, we found that pulse $P_6$ was clearly
right-handed.

In order to better distinguish reversal and rotation, we propose
more intensive observations of the Phase relationship between
magnetic field intensity and pulsed radiation in TVLM 513. The main
goals of this program would be to determine whether the magnetic
field intensity really goes through zero at point A between $P_1$
and $P_2$, and to improve the surrounding polarization measurements
(Fig.~\ref{Fig3}). Especially, if the line-of-sight and transverse
magnetic field intensities are both observed to go through zero at
point A, we can conclude that magnetic field reversal is indeed
taking place and producing the pulsed radio signal.

Another prediction of the MO model is that in the most of cases, the
largest pulses appears to the right of a zero point in the magnetic
field (e.g., P2 in Fig. 3), because the induced electromotive force
reaches its maximum by and large at the zero point of magnetic
field, the charged particles should reach their maximum speed
sometime after the zero point generally. This prediction can be
verified in cases where the pulses are narrow.

\subsection{Relationship between Radio Pulse Duty Cycle and the Magnetic Field Profile}

Some dwarf stars, for example 2MASS J00361617 + 1821104 \citep{H08},
produce radio pulses with a high duty cycle. In other stars, such as
TVLM 513, the duty cycle is very low. The duty cycle can also
provide some information on changes in the magnetic field.

According to the MO model, radio radiation is intrinsically pulsed.
The pulse's duty cycle depends on the profile of magnetic field. If
the field profile is close to a square wave, the duty cycle will be
low. If it is close to sinusoidal, the duty cycle will be high.

A rotating magnetic field will always have a sinusoidal profile and
there will be no the above relationship between duty cycle and the
magnetic field profile. Consequently, whether a star's magnetic
field rotates or reverses can be determined by measuring its duty
cycle and magnetic field profile synchronously.

\subsection{Relationship between Radio Pulse Polarization and the Magnetic Field}
As previously mentioned, the MO model assumes that the magnetic axis
and spin axis are always aligned. This implies certain relationships
between radio polarization and the spin inclination:
   \begin{enumerate}
      \item If the spin inclination is zero, circular polarization will be
detected.
      \item If the spin inclination is 90 degrees, linear polarization will
be detected.
      \item At other inclinations, elliptical polarization will be detected.

   \end{enumerate}
From the above relationships, the following can be derived:
   \begin{enumerate}
      \item If the observed radio pulse is 100\% circularly polarized, no
transverse magnetic field will be observed.
      \item If the observed radio pulse is 100\% linearly polarized, no
line-of-sight magnetic field will be observed.
   \end{enumerate}

As the above statements are only true under the MO model is correct,
verifying (or refuting) them can reliably determine whether the
magnetic field reverses (or rotates).

Because the measurement for both the transverse magnetic field and
the line-of-sight one depends on the measurement of polarized
spectral line, this method in nature is a measurement of the
relationship between radio polarization and line spectrum
polarization.

Note that Faraday rotation due to light passing through the
magnetosphere and various depolarization effects have been neglected
in this method. These factors may create a more complicated
relationship between radio pulse polarization and magnetic field. In
any case, the investigation on the relationship between radio signal
and line spectrum should provide new insights into the behavior of
stellar magnetic fields.

\section{Discussion}

Babcock (1958) found that all typical $\alpha$-variables exhibit a
fair degree of symmetry about the \emph{\textbf{H}} = 0 axis in
their reversals. That is to say, the polarity of all of the magnetic
fields reverses symmetrically. It is very difficult to explain this
result using the oblique rotator model. Up to now, Babcock's result
constitutes the most convincing evidence for the solar cycle model.
The observation that radio pulses from CU Virginis appear close to
the magnetic field null also favors the MO model, which is based on
the solar cycle model. Although both results support the magnetic
field reversal interpretation, the magnetic field reversal model is
not acceptable to many people. The main reason is that the turbulent
dynamo model cannot achieve periods under the order of a few days
--- to say nothing of the millisecond cycles typical of pulsars.

We put forth that the turbulent dynamo model is still unsupported,
and thus should not be taken as evidence for against magnetic field
reversals. Rather, more observing methods should be developed to
determine whether stellar magnetic fields reverse or rotate. If all
the observations described in this paper support a magnetic field
reversal, the present turbulent dynamo model will have to be
modified (or even ruled out).

Noteworthily, some results of data processing, such as \citet{M08}
and \citet{S08}, seem to be evidence to support magnetic field
rotation. In fact, these results need on the contrary to be
supported by the magnetic field rotation model. If magnetic field
reverses, and rotation model is wrong , all of the data processing
to convert time-domain signals into the three-dimensional image data
are only pure mathematical manipulations, without any really
physical significance. It is entirely possible that the reversal and
oscillation of magnetic field can influence the apparent elemental
abundance of star and the characteristics of spectral line and then
the periodical time-domain signals emerge.

If the magnetic field reverses, then most periodic stellar phenomena
($v$ sin $i$, H$\alpha$, apparent elemental abundance, luminosity,
pulse radiation, etc.) should vary on the magnetic cycle and bear no
direct relation to the rotation period. (Note, however, that the
star's rotation rate exerts a direct influence on its internal
dynamics, and may therefore affect the period of magnetic field
reversal). If observations of dwarf stars can authenticate reversal
periods on the order of hours, the prevailing model of magnetic
field generation will have to be changed. Its replacement (for
example, the unipolar dynamo model) may be able to explain the very
short reversal periods, from hundreds of seconds to milliseconds,
observed in white dwarfs and pulsars.

If believing that the sun is a common dwarf star, it should be
believed that most of dwarf stars in the sky have alternating
magnetic field which is similar to the sun's magnetic field. If a
quasi-stationary magnetic field with a very big magnetic dip angle
really exists in some stars, it means that there are two kinds of
magnetic fields, the alternating magnetic field and the
quasi-stationary one. The questions will be followed. Which are
alternating ones or quasi-stationary ones? What conditions can
decide the type of magnetic field in a star? What is the low limit
of periods for the sun-like magnetic field? In order to explain
these questions, some reliable methods should be found to
distinguish one type of the magnetic field from another. This paper
showed just an attempt to find such methods.

\clearpage

\end{document}